\documentstyle[11pt]{article}

\def\AFOUR{%
\setlength{\textheight}{8.0in}%
\setlength{\textwidth}{5.75in}%
\setlength{\topmargin}{-0.375in}%
\hoffset=-.5in%
\renewcommand{\baselinestretch}{1.17}%
\setlength{\parskip}{6pt plus 2pt}%
}

\AFOUR                                           

\expandafter\ifx\csname amssym.def\endcsname\relax \else\endinput\fi
\expandafter\edef\csname amssym.def\endcsname{%
       \catcode`\noexpand\@=\the\catcode`\@\space}
\catcode`\@=11
\def\undefine#1{\let#1\undefined}
\def\newsymbol#1#2#3#4#5{\let\next@\relax
 \ifnum#2=\@ne\let\next@\msafam@\else
 \ifnum#2=\tw@\let\next@\msbfam@\fi\fi
 \mathchardef#1="#3\next@#4#5}
\def\mathhexbox@#1#2#3{\relax
 \ifmmode\mathpalette{}{\m@th\mathchar"#1#2#3}%
 \else\leavevmode\hbox{$\m@th\mathchar"#1#2#3$}\fi}
\def\hexnumber@#1{\ifcase#1 0\or 1\or 2\or 3\or 4\or 5\or 6\or 7\or 8\or
 9\or A\or B\or C\or D\or E\or F\fi}

\font\tenmsa=msam10
\font\sevenmsa=msam7
\font\fivemsa=msam5
\newfam\msafam
\textfont\msafam=\tenmsa
\scriptfont\msafam=\sevenmsa
\scriptscriptfont\msafam=\fivemsa
\edef\msafam@{\hexnumber@\msafam}
\mathchardef\dabar@"0\msafam@39
\def\dashrightarrow{\mathrel{\dabar@\dabar@\mathchar"0\msafam@4B}}
\def\dashleftarrow{\mathrel{\mathchar"0\msafam@4C\dabar@\dabar@}}

\def\ulcorner{\delimiter"4\msafam@70\msafam@70 }
\def\urcorner{\delimiter"5\msafam@71\msafam@71 }
\def\llcorner{\delimiter"4\msafam@78\msafam@78 }
\def\lrcorner{\delimiter"5\msafam@79\msafam@79 }
\def\yen{{\mathhexbox@\msafam@55}}
\def\checkmark{{\mathhexbox@\msafam@58}}
\def\circledR{{\mathhexbox@\msafam@72}}
\def\maltese{{\mathhexbox@\msafam@7A}}
\def\circledS{{\mathhexbox@\msafam@73}}

\font\tenmsb=msbm10
\font\sevenmsb=msbm7
\font\fivemsb=msbm5
\newfam\msbfam
\textfont\msbfam=\tenmsb
\scriptfont\msbfam=\sevenmsb
\scriptscriptfont\msbfam=\fivemsb
\edef\msbfam@{\hexnumber@\msbfam}
\def\Bbb#1{{\fam\msbfam\relax#1}}
\def\widehat#1{\setbox\z@\hbox{$\m@th#1$}%
 \ifdim\wd\z@>\tw@ em\mathaccent"0\msbfam@5B{#1}%
 \else\mathaccent"0362{#1}\fi}
\def\widetilde#1{\setbox\z@\hbox{$\m@th#1$}%
 \ifdim\wd\z@>\tw@ em\mathaccent"0\msbfam@5D{#1}%
 \else\mathaccent"0365{#1}\fi}
\font\teneufm=eufm10
\font\seveneufm=eufm7
\font\fiveeufm=eufm5
\newfam\eufmfam
\textfont\eufmfam=\teneufm
\scriptfont\eufmfam=\seveneufm
\scriptscriptfont\eufmfam=\fiveeufm

\csname amssym.def\endcsname

\makeatletter
\def\section{\@startsection {section}{1}{\z@}{-3.5ex plus -1ex minus 
 -.2ex}{2.3ex plus .2ex}{\large\sc}}
\def\subsection{\@startsection{subsection}{2}{\z@}{-3.25ex plus -1ex minus 
 -.2ex}{1.5ex plus .2ex}{\normalsize\sc}}
\makeatother

\makeatletter
\@addtoreset{equation}{section}                        

\makeatother

\newcommand{\nc}{\newcommand}
\newcommand{\rnc}{\renewcommand}

\nc{\be}{\begin{equation}}
\nc{\ee}{\end{equation}}
\nc{\bea}{\begin{eqnarray}}
\nc{\eea}{\end{eqnarray}}

\nc{\trac}[2]{{\textstyle\frac{#1}{#2}}}

\nc{\ex}[1]{\mbox{e}^{\,\textstyle#1}}

\nc{\CC}{\Bbb{C}}
\nc{\HH}{\Bbb{H}}
\nc{\PP}{\Bbb{P}}
\nc{\RR}{\Bbb{R}}
\nc{\ZZ}{\Bbb{Z}}
\nc{\II}{\Bbb{I}}
\nc{\EE}{\Bbb{E}}
\nc{\SS}{\Bbb{S}}

\rnc{\a}{\alpha}
\nc{\ab}{\alpha^{*}}
\nc{\al}{\a^{l}}
\rnc{\d}{\delta}
\nc{\ga}{\gamma}
\nc{\la}{\lambda}
\nc{\lal}{\la_{l}}
\nc{\f}{\phi}
\nc{\fb}{\bar{\phi}}
\nc{\p}{\psi}

\nc{\e}{\varepsilon}
\nc{\eb}{\bar{\e}}
\rnc{\c}{\chi}
\nc{\cb}{\bar{\chi}}
\nc{\eps}{\epsilon}
\rnc{\t}{\theta}
\nc{\tb}{\bar{\theta}}
\nc{\om}{\omega}

\rnc{\P}{\Psi}
\nc{\pl}{\P_{L}}
\nc{\pdr}{\P^{\dag}_{R}}
\nc{\G}{\Gamma}
\nc{\Ga}{\Gamma}

\nc{\sig}{\sigma}
\nc{\sk}{\sigma_{k}}
\nc{\sa}{\sigma_{a}}
\nc{\Bb}{\bar{B}}

\nc{\symx}{\circledS}

\nc{\Q}{\bar{Q}}
\nc{\C}{{\cal A}/{\cal G}}                
\nc{\A}[1]{{\cal A}^{#1}/{\cal G}^{#1}}  
\nc{\RC}{{\cal R}_{\C}}                 
\nc{\RM}{{\cal R}_{\M}}                
\nc{\RX}{{\cal R}_{X}}
\nc{\RY}{{\cal R}_{Y}}

\nc{\ad}{\mathop{\mbox{ad}}\nolimits}
\nc{\tr}{\mathop{\mbox{tr}}\nolimits}
\nc{\Tr}{\mathop{\mbox{Tr}}\nolimits}
\nc{\Det}{\mathop{\mbox{Det}}\nolimits}
\rnc{\det}{\mathop{\mbox{det}}\nolimits}
\nc{\rk}{\mathop{\mbox{rk}}\nolimits}
\nc{\diag}{\mbox{diag}}
\nc{\ra}{\rightarrow}
\nc{\Ra}{\Rightarrow}
\nc{\LRa}{\Leftrightarrow}
\nc{\lra}{\leftrightarrow}
\nc{\ot}{\otimes}
\rnc{\ss}{\subset}
\nc{\ul}{\underline}
\nc{\nul}{\noindent\underline}
\nc{\non}{\nonumber\\}
\rnc{\S}{\Sigma}
\nc{\tp}{2\pi i}
\nc{\del}{\partial}
\nc{\dbar}{\bar{\del}}
\nc{\dx}{\dot{x}}
\nc{\zb}{\bar{z}}

\rnc{\lg}{\log g^{2}}
\nc{\lv}{\log V_{s}}
\nc{\vs}{V_{s}}
\rnc{\ln}{\log \N}
\nc{\ls}{\ell_{s}}
\nc{\N}{{\cal N}}
\nc{\M}{{\cal M}}
\nc{\F}{{\cal F}}
\nc{\E}{{\cal E}}
\rnc{\P}{{\cal P}}
\nc{\I}{{\cal I}}
\nc{\IIt}{$\widetilde{\mbox{II}}$}
\nc{\gst}{\widetilde{g_{s}}}
\nc{\gsh}{\widehat{g_{s}}}
\nc{\lsh}{\widehat{\ls}}
\nc{\rllh}{\widehat{R_{11}}}
\nc{\lph}{\widehat{\ell_{P}}}
\nc{\mnd}{M_{Nd}}

\nc{\nam}{\nabla_{\mu}}
\nc{\nan}{\nabla_{\nu}}

\nc{\mat}[4]{\left(\begin{array}{cc}#1&#2\\#3&#4\end{array}\right)}
 
\nc{\r}[1]{\mathbf{#1}}
\nc{\rb}[1]{\overline{\mathbf{#1}}}

\nc{\gi}{\gamma_{i}}
\nc{\gj}{\gamma_{j}}

\nc{\subs}[1]{{\vspace*{0.5cm}}%
{\noindent\underline{\small\sc #1}}
{\vspace*{0.3cm}}}

\nc{\chap}[1]{{\clearpage}%
\begin{center}%
{\noindent\underline{\large\sc #1}}{\addcontentsline{toc}{section}{#1}}%
\end{center}%
{\vspace*{0.3cm}}}

\newcommand{\ba}{\begin{eqnarray}}
\newcommand{\ea}{\end{eqnarray}}

\def\slash#1{\setbox0=\hbox{$#1$}#1\hskip-\wd0\hbox to\wd0{\hss\sl/\/\hss}}

\begin{document}
\global\parskip=4pt

\makeatletter
\begin{titlepage}

\vskip .5in
\begin{center}
{\Large\sc On the Relationship between the Rozansky-Witten\\[.2in]
           and the 3-Dimensional Seiberg-Witten~Invariants}\\
\vskip 0.7in
{\large\sc Matthias Blau}\footnote{{\tt e-mail: mblau@ictp.trieste.it}} {\sc
and} {\large\sc George Thompson}\footnote{{\tt e-mail:
thompson@ictp.trieste.it}}
\vskip .1in
The Abdus Salam ICTP \\
P.O. Box 586 \\
34100 Trieste \\
Italy\\

\end{center}
\vskip .4in
\begin{abstract}
\noindent 
The Seiberg-Witten analysis of the low-energy effective action of
$d=4$ $N=2$ SYM theories reveals the relation between the Donaldson and
Seiberg-Witten (SW) monopole invariants. Here we apply analogous reasoning
to $d=3$ $N=4$ theories and propose a general relationship between
Rozansky-Witten (RW) and 3-dimensional Abelian monopole invariants.
In particular, we deduce the equality of the $SU(2)$ Casson invariant and
the 3-dimensional SW invariant (this includes a special case of
the Meng-Taubes theorem relating the SW invariant to Milnor torsion).
Since there are only a finite number of basic RW invariants of a given
degree, many different topological field theories can be used to represent
essentially the same topological invariant. 
This leads us to advocate
using higher rank Abelian gauge theories to shed light on the higher
(non-Abelian) RW invariants and we write down candidate higher rank SW
equations.
\end{abstract}

\end{titlepage}
\makeatother

\begin{small}
\tableofcontents
\end{small}

\setcounter{footnote}{0}

\section{Introduction}

Quite a few years ago Witten showed how Donaldson theory finds a natural
place in the context of quantum field theory \cite{ewdon}. Formal
developments soon uncovered the position of other theories in this
pantheon (see e.g.\ \cite{bbrt} for a review).

After Seiberg and Witten solved the $N=2$ supersymmetric Yang-Mills theory
in 4 dimensions \cite{sw1} the full force of quantum field theory
techniques was brought to bear on various mathematical theories. Perhaps
the most celebrated result 
is the (still to be fully established) equivalence
of the 4-dimensional Seiberg-Witten monopole invariants \cite{ewsw}
and the Donaldson invariants. 

The purpose of this paper is to apply analogous reasoning to
3-dimensional supersymmetric gauge theories and their associated
invariants. The relationship between the 4- and the 3-dimensional
theories has been analysed in great detail in \cite{mm}. We stay
firmly in 3 dimensions. 

One of our motivations was to understand a theorem of Meng and Taubes
from a physics point of view. Their theorem equates the Seiberg-Witten
invariants of a 3-manifold $M$ with its Milnor torsion, providing
$b_{1}(M)\geq 1$ \cite{mt}. On the other hand it is known that the Casson
invariant with the same homological constraint on $M$ is given in terms
of the Alexander polynomial of $M$ \cite{les}. The relationship between
Milnor torsion and the Alexander polynomial \cite{tu} means that the
Seiberg-Witten invariant equals the Casson invariant (for $b_{1}\geq
1$). Why should this be so? Amongst other things we answer this question.

We focus on $N=4$ supersymmetric Yang-Mills theories with 
gauge group $G$ and combine the following well known facts about these
theories:

\begin{enumerate}
\item The Coulomb branch of any $N=4$ SYM theory in 3 dimensions
is a (possibly singular) hyper-K\"ahler manifold \cite{sw3}.
\item Topologically twisting the low-energy description of this theory
one obtains the perturbative Rozansky-Witten \cite{rw} (or generalized
Casson) invariants.
\item Topologically twisting the microscopic theory instead, 
one obtains a topological field theory with 
two topological charges, $N_{T}=2$, 
modelling the de Rham complex and formally
calculating an Euler characteristic of some gauge theory moduli space
\cite{btn21,cvew,dm,btn22}.
\end{enumerate}

The two topological field theories, the one calculating the Euler
characteristic and the other
yielding perturbative invariants, arise on twists of the same physical
theory albeit at different energy scales. Since the topological theory
should not care what scale one is working at one immediately arrives
at the equality of the two types of topological field theories. This is
the same reasoning that one uses to establish that the Seiberg-Witten
invariants on a 4-manifold are equivalent to the Donaldson
polynomial invariants. 

The paradigm then is that the topological invariants that arise from
twisting the original gauge theory match the topological invariants that
arise on twisting the low energy effective theory.

In 3 dimensions, this reasoning leads to non-trivial results even in the
case that the gauge group of the microscopic theory is Abelian. Indeed,
as we will see below, applying this reasoning to the $N=4$ $U(1)$
theory with $N_{f}=1$ hypermultiplet yields the equivalence of the
3-dimensional version of the Seiberg-Witten monopole invariants and
the Casson (actually Casson-Lescop-Walker) invariant, not just for
$b_{1}\geq 1$ (this is part of the content of the Meng-Taubes theorem)
but also for the (mathematically much more subtle) case $b_{1}=0$.

But this is just the first in a whole hierarchy of observations one can
deduce in this way. The crucial additional input is the fact that,
denoting the hyper-K\"ahler manifold apapearing as the target space
of the Rozanzky-Witten sigma model by $X$, $\dim_{{\Bbb R}}X=4n$,
\begin{enumerate}
\addtocounter{enumi}{3}
\item 
there are only a {\em finite} number of independent
perturbative\footnote{In the following we refer to the Rozansky-Witten
invariants as perturbative or finite type invariants. However, that the
RW invariants are of finite type has not
been rigorously established even though there is a lot of evidence in
favour of this. All we really need is that
there are a finite number of RW invariants at any order in
perturbation theory and this is obviously true.} invariants 
for each $n$. For example the number of independent perturbative
invariants for $n=1,2,3,4$, are $1,1,1,2$. 
\end{enumerate}
Combining these facts one realises that many (possibly infinitely
many) different gauge theoretic moduli spaces have Euler
characteristics which (as functions of the 3-manifold) are linearly
dependent. 

The main mystery of the perturbative invariants is their relationship
to data of the 3-manifold. The Rozansky-Witten viewpoint provides a
resolution to this dilemma as it relates generalized Casson invariants
to the perturbative invariants. The Casson invariants are tied to
$\pi_{1}(M)$ and so the finite type invariants must also know about
the fundamental group. However, our point of view is that the
Rozansky-Witten approach offers a plethora of different moduli space
interpretations of the finite type invariants. Ultimately, it may well
be that the most interesting information is the relationship to the
fundamental group of $M$. Nevertheless, the availability of the many
different representations of the invariants offers computational
power. One general conclusion is that it is worthwhile to try to
simplify the situation as much as possible. The Seiberg-Witten
invariants on a 4-manifold are analytically more tractable than the
Donaldson invariants. Likewise, rather than studying the $SU(n)$ Casson
invariants directly, it may well be profitable to look at the associated
Abelian Seiberg-Witten equations. We take a preliminary look at such
equations in section 3.2.

There remains a mystery, however. Why do all the gauge theoretic
moduli spaces lead to perturbative, presumably finite type, invariants?

\section{The Casson, RW and SW Invariants}

In this section we make use of the identification of the various
topological field theories to relate their associated topological
invariants. We concentrate on those theories which have as their
Coulomb branch moduli space some 4-dimensional hyper-K\"{a}hler
space. Given a
hyper-K\"{a}hler manifold $X$ of real dimension 4 the associated
Rozansky-Witten invariant for a 3-manifold $M$ is, in the notation of
\cite{ht}, 
\be
Z_{X}^{RW}[M] = {\mathbf e}(X) \, \lambda(M), \label{1}
\ee
where ${\mathbf e}(X)$ is the integral of the Euler
class\footnote{Note that for $X$ non-compact this is not
necessarily the Euler characteristic.} of $X$ and
$\lambda(M)$ is the suitably normalised
Casson-Lescop-Walker invariant \cite{walk},
\cite{les} (which extends the Casson invariant, originally
defined for integral homology spheres, to all 3-manifolds).

By the discussion in the Introduction we learn that any $N=4$
theory with a 4-dimensional Coulomb branch yields (upon twisting)
a topological invariant proportional to the Casson invariant. This
includes gauge theories with group $U(1)$ and any number of charged
hypermultiplets or the gauge group $SU(2)$ and any number of fundamental
hypermultiplets. In the list of such theories one possibly also has
examples with compact Coulomb branches, obtained via toroidal dimensional
reduction \cite{int}. As the only perturbative invariant available is
the Casson invariant, the Rozansky-Witten invariants associated to all
of these theories are proportional to the Casson invariant.

In turn this means that the
wildly different moduli spaces associated with the gauge theories all
have Euler characteristic proportional to the Casson invariant, the
proportionality factor being ${\mathbf e}(X)$. For example the $U(1)$
gauge theory with one charged hypermultiplet, when
twisted, yields the topological field theory corresponding to the
3-dimensional Seiberg-Witten equations. The pure $SU(2)$ theory
yields the Casson invariant \cite{rw}, while the $SU(2)$
theory with one hypermultiplet is the topological field theory for the
$SU(2)$ Seiberg-Witten monopole moduli space. We take a look at all of
these spaces next.

\subsection{The Casson and the RW Invariants}

The starting physical theory is the pure $N=4$ supersymmetric $SU(2)$
gauge theory. It has been argued that this theory, when twisted,
calculates the Casson invariant (this essentially goes back to Taubes'
gauge theoretic interpretation of the Casson invariant \cite{t} which
was given a topological field theory interpretation in \cite{ewtca}
subsequently elaborated upon in \cite{btn21}).
On the other hand the low energy
effective theory has as its moduli space the Atiyah-Hitchin manifold
$X_{AH}$ \cite{sw3},
which is the $SU(2)$ 2-monopole moduli space. It is known that
\be
{\mathbf e}(X_{AH}) = 1,
\ee
and so the Rozansky-Witten (\ref{1}) invariant in this case really is
equal to the Casson invariant, as it should be!

\subsection{The Seiberg-Witten and the RW Invariants}

Consider a $N=4$, $U(1)$, supersymmetric gauge theory with $N_{f}=1$
hypermultiplets. Seiberg and Witten have shown that the moduli space
is the Taub-Nut hyper-K\"{a}hler manifold $X_{TN}$ \cite{sw3}. 
If we now twist this theory
we obtain a supersymmetric sigma model with target
space $X_{TN}$. The integral of the Euler class of the
Taub-Nut is
\be
{\mathrm e}(X_{TN}) = 1.
\ee
The Rozansky-Witten invariant is therefore once more equal to the Casson
invariant. 

On the other hand, as we will now see, the topological invariant
associated with the microscopic theory counts solutions to the
3-dimensional Abelian Seiberg-Witten monopole equations.

The gauge theory setting is that of $N=4$ supersymmetric Yang-Mills
theory with gauge group $U(1)$ equipped with a charged hypermultiplet.
The $N=4$ theory is most usefully regarded as the dimensional reduction
of the six-dimensional $N=1$ theory to 3 dimensions. This
exhibits the R-symmetry group $SU(2)_{R}\times SU(2)_{N}$ of the
3-dimensional theory, $SU(2)_{R}$ being the R-symmetry group of
the six-dimensional theory and $SU(2)_{N}$ the rotation group in the
`internal' 3 dimensions.

The fields in
the charged hypermultiplet transform as
\be
{\mathrm Bosons}: \,\, (2,1,1)^{\pm}, \;\;\;\; {\mathrm Fermions}: \,\,
(1,2,2)^{\pm} , \label{stats}
\ee
under $SU(2)_{R}\times SU(2)_{N}\times SU(2)_{E} \times U(1)$ where
$SU(2)_{E}$ is the space-time Lorentz group (for a more careful discussion 
of R-symmetry groups in the Euclidean versus Lorentzian theories and their
twists see \cite{btsym} - we will not have to worry about these issues here).

Now what
theory do we have if we twist the model directly without passing
to the low energy theory? Twisting in this case means that we consider
the diagonal, $SU(2)_{E'}$ of $SU(2)_{R} \times SU(2)_{E}$ to be the
new Lorentz group. From (\ref{stats}) we see that after twisting the
field content transforms as
\be
{\mathrm Bosons}: \,\, (1,2)^{\pm}, \;\;\;\; {\mathrm Fermions}: \,\,
(2,2)^{\pm} , \label{stats2}
\ee
under $SU(2)_{N}\times SU(2)_{E'} \times U(1)$. This, together with
the twisted vector multiplet, is precisely the field
content of the topological theory corresponding to the 3-dimensional
Seiberg-Witten equations 
and twisting the supersymmetric action leads us directly to the action for 
the topological theory 
\cite{cmwz,gt,ohta}.

Denoting the gauge field by $A$, its field
strength by $F$, and the (commuting) spinor field arising from the 
twisted hypermultiplet by $M$, the 3-dimensional SW equations are
\bea
F_{\mu\nu} & = & 
-\trac{i}{2}\overline{M} \sigma_{\mu\nu}M \\
\slash{D}(A) M & = & 0\;\;.
\eea
(We will occasionally write the first equation more compactly as
$F_{A} = * \overline{M}\gamma M$.)

In this way we have established that the physics of $N=4$ gauge theories
in 3 dimensions predicts
\be
SW(M) = \lambda(M).
\ee

\subsection{Evidence in Favour of SW = Casson}

There is abundant evidence in the mathematics literature that supports
this claim. Meng and Taubes have shown that for $b_{1}(M) \geq 1$
\be
\underline{{\mathrm SW}}(M, t_{i}) = \tau(M,t_{i})
\ee
where $\tau(M, t_{i})$ is the Milnor torsion of $M$ while the SW series
$\underline{{\mathrm SW}}(M, t_{i})$ of $M$ is defined e.g.\ in 
\cite{mt,mm}.
As a special case, the only one we will actually need, we have
\be
{\mathrm SW}(M) = \tau(M,1),
\ee
where ${\mathrm SW}(M)= \underline{{\mathrm SW}}(M, t_{i}=1)$ and $\tau(M,1) =
\tau(M, t_{i}=1)$. For $b_{1}(M)>1$ one has 
\be
\tau(M, t_{i}) = \Delta_{M}(t_{i}), \label{t2}
\ee
where $\Delta_{M}(t_{i})$ is the Alexander polynomial of $M$, symmetrized
in $t$ and $t^{-1}$. Lescop \cite{les} has shown for $b_{1}(M) >1$ that
$\lambda(M) = \Delta_{M}(t_{i}=1)$. Consequently, for $M$ such that
$b_{1}(M)>1$ \be {\mathrm SW}(M) = \lambda(M).  \ee

For $b_{1}(M)=1$ the relationship between the Milnor Torsion and the
Alexander polynomial is \cite{tu}
\be
\tau(M, t) = \frac{t \, \Delta_{M}(t)}{(1- t)^{2}} .\label{t1}
\ee
The right hand side expanded about $t=1$,
\be
\frac{t}{(1-t)^{2}}\Delta_{M}(1) + \frac{t}{2} \Delta_{M}^{(2)}(1) + \dots
\ee
is singular as $t \rightarrow 1$ and so must be suitably interpreted. 
Note that
\be
\frac{1}{(1-t)^{2}} =
\frac{d}{dt}\frac{1}{(1-t)}
=\frac{d}{dt}\sum_{n=0}^{\infty}t^{n}= \sum_{n=1}^{\infty}n t^{n-1}
\ee
which, as $t\rightarrow 1$, goes over to
\be
\lim_{s \rightarrow -1}\sum_{n}n^{-s} = \zeta(-1) = - \frac{1}{12} .
\ee
The regularised form of the limit is then
\be
\tau(M, 1) = - \frac{1}{12}\Delta_{M}(1) + \frac{1}{2}\Delta_{M}^{(2)}(1),
\ee
which agrees, once more, with the result found by Lescop for the
Casson invariant.

The most important case is when $b_{1}(M)=0$, i.e.\ when
$M$ is a rational homology 3-sphere. It has been shown that
for integral homology spheres the SW
invariant equals the Casson invariant \cite{lim}. 
This is conjectured to be the
case also for rational homology spheres and there is considerable
evidence for this \cite{n}. 

Actually there is a subtlety here. The Seiberg-Witten
equations are defined with some perturbation. For
$M$ a rational homology sphere one deforms the equations to
\be
F_{A} = * \overline{M}\gamma M + d \nu
\ee
where $\nu$ is a 1-form on $M$. With this choice of perturbation there is
a choice of metric such that the reducible solution set $(A,M)= (\nu, 0)$
is isolated from the irreducible solutions (which themselves form a finite
set). The problem is that this split is not so clean. As one varies the
metric and the perturbation, some of the irreducible points may collide
with the reducible solution, or some irreducible points may bubble off
from the reducible solution. This means that the count of the signed
sum of the irreducible points is not an invariant. On the other hand the
`total' path integral is formally metric and perturbation invariant. The
problem seems to have arisen because we have `excised' the reducible.

So the path integral may well be defined to count the
reducible point (in some fashion, depending on the metric and on the
perturbation) together with a signed sum of the irreducible
points. The usual BRST argument suggests that one can change the
metric and perturbation at will, providing the solution set remains
finite. In this case extra solutions to the Dirac equation may appear
at the reducible as one varies the parameters and a correction term is
required
to compensate the dissapearance of solutions away from the reducible.

There is a choice for the contribution of the reducible connection
in terms of $\eta$-invariants (see e.g.\ \cite{chen,lim,m}), 
\be
\frac{1}{2}\eta(\slash{D}_{\nu}) + \frac{1}{8}\eta (*d - d*)
\ee
which is known to ensure that the total contribution yields the Casson
invariant for integral homology 3-spheres. The same contribution is
conjectured to be correct for rational homology 3-spheres as well and
there is some evidence to support this \cite{n}. Note incidentally that
for rational homology spheres admitting a metric of positive scalar
curvature (e.g.\ Lens spaces) the above correction term is the only
contribution to the SW invariants as there are no non-trivial solutions to 
the SW monopole equations in this case. 

It is interesting to note that the Rozansky-Witten invariant is
affected by an analogous `correction' term. In this case the ambiguity
amounts to a choice of 2-framing of $M$. That choice must be fixed for
the invariant to coincide with $\lambda(M)$.

\subsection{The SU(2) Theory with Hypermultiplets}

Seiberg and Witten also studied the $SU(2)$ supersymmetric theory with
$N_{f}$ hypermultiplets in the fundamental representation \cite{sw3}. 
The Coulomb
branch moduli space depends on $N_{f}$. They give quite a complete
description of the spaces involved:
\begin{enumerate}
\item For $N_{f}=1$ the space is $\tilde{X}_{AH}$, the double cover of
$X_{AH}$.
\item For $N_{f} = 2$ the space is topologically and metrically $\left({\Bbb
R}^{3} \times S^{1}\right)/{\Bbb Z}_{2}$.
\item For $N_{f}>2$ they are ALE spaces with a $D_{N_{f}}$
singularity, ${\Bbb C}^{2}/\Gamma_{N_{f}-2}$.
\end{enumerate}
For $N_{f}\geq 2$ one should perhaps resolve the singularities and this
can be achieved by adding bare mass terms to the hypermultiplets. The
corresponding Rozansky-Witten invariants are
\begin{enumerate}
\item $N_{f}=1$: $2 \lambda(M)$ .
\item $N_{f}= 2$:  $0$ .
\item $N_{f}> 2$: $\left(2N_{f}-1 - \frac{1}{4(N_{f}-2)} \right)
\lambda(M)$. 
\end{enumerate}

We now need to see what moduli space one gets on the gauge theory
side. The reader will not be too surprised to learn that the
moduli space of the topological gauge theory is that of non-Abelian
monopoles. Non-Abelian monopole equations have been studied in 4
dimensions \cite{lm} and one may just as well consider them in 3
dimensions. Let ${\mathcal S}$ denote the spin bundle and $E_{i}$
be complex vector bundles associated to a principal $G$-bundle via
the representations $R_{i}$. Let 
$M_{i}$ be sections of ${\mathcal S} \otimes E_{i}$. 
The monopole equations are
\be
F^{a}_{A} = * \sum_{i = 1}^{N_{f}} \overline{M}_{i}\gamma 
T^{a}_{i}M_{i} , \;\;\;
\slash{D}_{A}M_{i} = 0, \label{nab}
\ee
where $T^{a}_{i}$ are the generators of the Lie algebra of $G$ in the
representation $R_{i}$.
$G=SU(2)$, ${\mathrm rank}(E)= 2$ and $N_{f}=1$ is the case
most studied in 4 dimensions. The situation with bare mass terms for the
hypermultiplet included has also been considered in that context.

The claim now is that the Euler characteristic of the non-Abelian
$SU(2)$ monopole equations (\ref{nab}) (suitably perturbed) equals the
multiple of the Casson invariant listed above.

\section{Higher Rank Seiberg-Witten Invariants}

We have, so far, only concentrated on theories which are proportional
to the usual Casson invariant or, equivalently, to those which have a
4-dimensional Coulomb branch moduli space. Considering a group of
larger rank or adding matter content in other representations (for
example the adjoint representation) leads to a higher dimensional
hyper-K\"{a}hler space. From the Rozansky-Witten side this means that
the invariant being probed is a higher order RW invariant. 

In order to see the higher order RW invariants certain
integrals of products of the Riemann curvature tensor on the
hyper-K\"{a}hler space must be non-zero. Roughly one has a dependence
of the type (see equation (10.17) in \cite{ht}) 
\be
\lambda^{k}_{X}(M) = B_{i}(X) \lambda^{k}_{i}(M) + \ldots  ,\label{rw}
\ee
where $\lambda^{k}_{i}(M)$ is the 3-manifold dependence of the $i$-th
RW invariant of order $k$, $B_{i}(X)$ is the dependence on the 
hyper-K\"{a}hler
manifold (it also depends on $b_{1}(M)$) and the ellipses denote
dependence on products of lower order RW invariants.  For $b_{1}(M) >
1$ one only needs ${\mathbf e}(X) \neq 0$
\be
\lambda^{k}_{X}(M)= {\mathbf e}(X) \, \left(\lambda(M)\right)^{k} \label{rw23}
\ee
and the ellipses in (\ref{rw}) are zero. For $b_{1}(M)=1$ one finds
that the invariant is \cite{ht}
\be
\lambda^{k}_{X}(M)= \int_{X} \, \hat{{\mathbf A}}(X)\, 
\prod_{i=1}^{n} \Delta_{M}
(e^{x_{i}}),\label{rw1} 
\ee
where the $x_{i}$ are the eigenvalues of the curvature 2-form of the
holomorphic tangent bundle. Expanding the $\hat{{\mathbf A}}$-genus 
and $\Delta_{M}$ in terms of the $x_{i}$ and keeping only the 
top-form components, one finds an expression of the form (\ref{rw}).
The dependence on the 3-manifold is
through classical invariants and the dependence on the
hyper-K\"{a}hler manifold is through characteristic classes.

The case of $b_{1}(M)=0$ is quite different. The
dependence on the hyper-K\"{a}hler manifold is not just through
characteristic classes of the holomorphic tangent bundle \cite{ht} but has a
rather more subtle dependence on the curvature tensor \cite{hs}. Also
the dependence on the 3-manifold is not through classical invariants,
which is just as well as otherwise there would be nothing new here.

The upshot is that one will have to make judicious choices of the
content of the gauge theory to `see' the higher order invariants. It
is believed that the pure $SU(n)$, $N=4$ theories have Coulomb
branches which probe some of these invariants. 

Incidentally these observations in a sense go both ways. If, for some
reason, the 3-manifold invariants in question are known, then one can 
use the above reasoning to obtain some information on the curvature
integrals $B_{i}(X)$ of hyper-K\"ahler manifolds instead. 

For example, `higher knowledge' from 4-dimensions suggests (at least
for large enough $b_{1}(M)$) that the Casson invariant should equal the
invariant that one obtains from the dimensional reduction of the theory
describing the 4-dimensional Seiberg-Witten equations. The reduced theory
is that for the 3-dimensional Seiberg-Witten equations. We would deduce,
therefore, that ${\mathbf e}(X_{TN}) =1$. That the reduced invariants come
out right has been shown by Mari\~{n}o and Moore \cite{mm}. Similarly the
reduction of the Donaldson theory itself must yield the Casson invariant
and so we find ${\mathbf e}(X_{AH})=1$. Happily both `predictions' hold
and the circle of ideas could have led us to predict that ${\mathbf
e}(X_{TN}) = {\mathbf e}(X_{AH})$.

Perhaps reasoning of this type in other settings, in particular for
$b_{1}(M) \geq 1$ where the RW invariants are classical 3-manifold
invariants, could be used to garner information about curvature integrals
on other hyper-K\"{a}hler manifolds.

\subsection{$b_{1}(M)\geq 1$ and Implications of the Theorem of Meng and Taubes}

We start by exploring the higher order RW invariants for 3-manifolds
with $b_{1}(M)\geq 1$.
We can rewrite the Rozansky-Witten expression for any $M$ with
$b_{1}(M) \geq 1$ in the following compact form
\be
\int_{X} \prod_{i=1}^{n}\, x_{i}^{2} \, \tau(M, e^{x_{i}}) \label{rw123}.
\ee
This makes sense: for $b_{1}(M) \geq 2$, the relationship
between the Milnor torsion and the Alexander polynomial (\ref{t2})
and the fact that $\Delta(M, t_{j})$ is regular as the $t_{j}
\rightarrow 1$ guarantees that (\ref{rw123}) becomes (\ref{rw23}),
while for $b_{1}(M)=1$ one makes use of (\ref{t1}) to show that
(\ref{rw123}) becomes (\ref{rw1}). 

Let us now consider some $N=4$ gauge theory and the topological
twist of this microscopic theory.
We can refer to all of the equations that one arrives at on the
topological gauge theory side as generalized Seiberg-Witten
equations. This includes equations that may not involve any matter
fields at all (as for the Casson invariant and its
generalizations). Let $SW^{G}$ denote the invariant obtained from
the generalized Seiberg-Witten equations (formally, as mentioned before,
this is the signed sum \cite{btn21,btn22} of Euler characteristics of the 
solution space). 

Now consider the low-energy description of this theory, in terms of 
a supersymmetric sigma model with target space the hyper-K\"ahler
Coulomb branch $X$. Equating the generalized SW invariant and the 
RW invariant we thus arrive at the

\underline{{\bf Conjecture}}: For a connected, compact, closed oriented
3-manifold, M, with $b_{1}(M) \geq 1$, the generalized Seiberg-Witten
invariant is
\be
{\mathrm SW}^{G}(M) = \int_{X} \prod_{i=1}^{n}\, x_{i}^{2} \, \tau(M,
e^{x_{i}})  . \label{mtg}
\ee
Put another way: The generalized Seiberg-Witten invariants are given
entirely in terms of Milnor Torsion. But by
the Meng-Taubes theorem this means that they are determined by 
the Abelian SW invariant \underline{SW}(t). 
Consequently, it appears that neither the
generalised SW nor the Casson type invariants shed new light on 3-manifolds
with $b_{1} \geq 1$.

\subsection{$b_{1}(M)=0$ and Abelian Higher Rank Seiberg-Witten Invariants}

We have seen above that the only real case of interest for the
types of invariants we have been considering is the notoriously
subtle case of rational homology spheres, $b_{1}(M)=0$. 
Since there is only one new perturbative invariant for $n=2$ (and also
for $n=3$) all theories with an 8 (12) dimensional Coulomb branch yield
essentially the same invariant (ignoring lower order invariants). The
3 (4) -monopole $SU(2)$ moduli space is believed to correspond to the
$SU(3)$ ($SU(4)$) Casson invariant. 

However, Abelian theories are usually easier to get a handle on than
non-Abelian ones. It is thus reasonable to try to probe the higher order
Rozansky-Witten invariants for rational homology spheres
with $N=4$ $U(1)^{r}$ gauge theories coupled to
various charged hypermultiplets. We will simply refer to the gauge
theoretic equations that the topological field theory leads to as
Abelian Seiberg-Witten equations. 

To lead to `useful' generalizations of the standard SW equations, these
theories should satisfy the following three conditions:
\begin{enumerate}
\item First of all, the resulting SW equations should not be equivalent to 
a set of $r$ decoupled ordinary SW equations, as in that case one would
only probe the already well known term $\sim {\mathbf e}(X) \lambda(M)^{r}$
in the expansion of the RW invariant.
\item Alternatively and equivalently, the matter content will have to be
chosen appropriately to ensure that the Coulomb branch moduli space has
a metric with some of the $B_{i}(X)$ (\ref{rw}) other than ${\mathbf e}(X)$
non-zero. 
\item
Finally, in order to have a well defined counting problem on the SW side,
one would like the moduli space to be compact. In practice this can be
established most readily if an analogue of the Weitzenb\"ock argument 
of \cite{ewsw} can be used to bound the norm of the spinors and the 
gauge field strenghts in terms of the scalar curvature of the 3-manifold.
This, together with an argument about reducible solutions, is sufficient
to establish compactness of the moduli space. 
\end{enumerate}

One Abelian Seiberg-Witten system that might be of interest is
that obtained from the twist of the $N=4$
supersymmetric $U(1)^{r}$ theory with $r$, appropriately charged,
massless hypermultiplets studied e.g.\ in \cite{si,hw}. The Coulomb
branch of these theories is a multi-dimensional version of the 
Taub-NUT metric first obtained in \cite{lwy} as a particular
$SU(r+2)$ monopole moduli space.

The corresponding SW equations are special cases of a more 
general system of equations for gauge group $U(1)^{r}$ coupled
to $r$ charged hypermultiplets. These equations are,
with $i,j = 1, \ldots, r$,
\bea
F^{i}_{\mu\nu} & = & 
-\trac{i}{2}\sum_{j}\left(\overline{M}_{j} 
\sigma_{\mu\nu}M_{j}\right) E_{ji}
\label{f}\\
\slash{D}(\sum_{j}E_{ij}A_{j} ) M_{i} & = & 0  \;\;\; ({\mathrm no\; sum
\; over}\; i)
\eea
where the $E_{ij}$ are the $j$-th charges of the $i$-th
monopole.  Under suitable conditions on the charge matrix $E_{ij}$ 
it is possible to establish bounds on the norms
of the $M_{i}$ and $F$ using the Weitzenb\"ock arguments of \cite{ewsw}. 
For example, for $r=2$ we find that a sufficient condition is 
\be
\det E \neq 0\;\;.
\ee
Moreover, it is still true that there are no non-trivial solutions
to the equations if $M$ admits a metric with scalar curvature $R>0$, 
just as for the $U(1)$ SW equations. To
establish that these equations are not equivalent to a pair of uncoupled
SW equations, and hence really probe the higher order RW invariants, it
is e.g.\ sufficient to show that the integral of $\Tr R_{X}^{4}$, $R_{X}$ the
Riemann curvature two-form of the Coulomb branch moduli space $X$, is non-zero.
We will describe these and other aspects of the problem in \cite{btsw}.

\subsubsection*{Acknowledgements}

We are grateful to M.S.\ Narasimhan for discussions.  This work was
supported in part by the EC under the TMR contract ERBFMRX-CT96-0090.

\rnc{\Large}{\normalsize}

\end{document}